# Structural and magnetic properties of core-shell Au/Fe$_3$O$_4$ nanoparticles

L. León Félix[1,2], J. A. H. Coaquira[1], M. A. R. Martínez[1], G. F. Goya[2], J. Mantilla[1], M. H. Sousa[3], L. de los Santos Valladares[4], C. H. W. Barnes[4] & P. C. Morais[1,5]

[1] Laboratory of Magnetic Characterization, Instituto de Física, Universidade de Brasília, DF 70910-900, Brasília, Brazil.

[2] Instituto de Nanociencia de Aragón (INA), Universidad de Zaragoza, 50018 Zaragoza, Spain.

[3] Green Nanotechnology Group, Faculdade de Ceilândia, Universidade de Brasília, Ceilândia, DF 72220-900, Brasília, Brazil.

[4] Cavendish Laboratory, Department of Physics, University of Cambridge, J.J Thomson Av., Cambridge CB3 0HE, United Kingdom

[5] School of Chemistry and Chemical Engineering, Anhui University, Hefei 230601, China

Correspondence and requests for materials should be addressed to L. de los Santos Valladares (ld301@cam.ac.uk)

We present a systematic study of core-shell Au/Fe$_3$O$_4$ nanoparticles produced by thermal decomposition under mild conditions. The morphology and crystal structure of the nanoparticles revealed the presence of Au core of $\langle d \rangle$ = (6.9±1.0) nm surrounded by Fe$_3$O$_4$ shell with a thickness of ~3.5 nm, epitaxially grown onto the Au core surface. The Au/Fe$_3$O$_4$ core-shell structure was demonstrated by high angle annular dark field scanning transmission electron microscopy analysis. The magnetite shell grown on top of the Au nanoparticle displayed a thermal blocking state at temperatures below $T_B$ = 59 K and a relaxed state well above $T_B$. Remarkably, an exchange bias effect was observed when cooling down the samples below room temperature under an external magnetic field. Moreover, the exchange bias field ($H_{EX}$) started to appear at $T$~40 K and its value increased by decreasing the temperature. This effect has been assigned to the interaction of spins located in the magnetically disordered regions (in the inner and outer surface of the Fe$_3$O$_4$ shell) and spins located in the ordered region of the Fe$_3$O$_4$ shell.

**Introduction**

Core-shell nanoparticles (CSNPs) have evolved as key materials because their applications as bifunctional nanomaterials. Different chemical and physical properties of the core-shell interface and their components contribute to the complexity of the system and attract efforts to understand them. In recent years different core-shell gold nanostructures have been studied regarding their catalytic properties[1,2], surface plasmon resonance[3], and other phenomena[4,5,6]. Metal-based magnetic core-shell nanoparticles would provide excellent magnetic sensitivity and biocompatibility. The well-established synthesis methods afford good reproducibility, narrow size distribution, and good control over the shape and size of the nanoparticles (NPs)[7]. The latest is important because many of the CSNPs' properties depend on the size, shape and composition. The collective behavior associated with the interparticle, dipolar and/or exchange interactions is critically important to understand the magnetic response of an ensemble of NPs. However, reports devoted to magnetostatic and magnetodynamic properties of core-shell structures' interface with diamagnetic cores and magnetic shells remain rather scarce.

In the case of NPs, the exchange-coupling between two different magnetic regions such as a disordered surface layer and a magnetically ordered core drives to interesting effects. Once the interface region between the core and shell layer contains broken bonds and lower coordination number that break down the translation symmetry, it generates randomness in the exchange interaction and drives to magnetic frustration. It is known that the effect of surface spins disorder becomes stronger as NPs[8,9,10] reduce size due to the large surface-to-volume ratio. Considering that NPs with high anisotropy exhibit weaker surface effects, they are expected to display reducing exchange bias effect as the size is further reduced[11], leading to an enhanced contribution to the complexity of the system's magnetic response. The exchange bias effect is reflected in a shift of the hysteresis loop during field-cooling processes[12,13]. Moreover, recent works have reported the occurrence of exchange bias effect in bicomponent systems, such as in Au/Fe$_3$O$_4$ CSNPs, which was related to the presence of a spin-glass-like layer at the particle surface[14]. However, in single-phased magnetic NPs this effect have also been attributed to disordered surface spins[9,15,16]. The magnetic response of multicomponent nanostructures depends on various parameter such as the atomic structure of the interface (uncompensated spins), finite-size effect, morphology, and surface effects[17,18]. The shell thickness or differences between effective anisotropies[19] are crucial factors, which determine the strength of the exchange bias effect in NPs. The temperature dependence of the exchange bias field can provide clues to determine whether the magnetic response results from a finite size effect or it is related to the presence of a surface layer with canted spins[10].

Reports about synthesis routes of NPs aiming to avoid agglomeration and polydispersed systems emphasize difficulties in the fabrication of new materials. We circumvented these drawbacks by employing the high temperature thermal decomposition method, which uses metallic precursors in the presence of organic surfactants[20]. This method is well established and widely used to produce highly crystalline, monodisperse size distribution, and shape-controlled NPs. Shape depends on the synthesis temperature, reaction atmosphere and other conditions such as availability of unsaturated bonds in the synthesis solvent. The presence of insulating capping ligands on the surface of magnetic NPs prevents formation of agglomerates via chains, leading to pure long-range (dipolar) interactions. It has been shown that while thermal decomposition processes are used to synthesize Fe$_{(1-x)}$O phases CO surface groups are generated, which actively participate in the reduction of Iron (III) and therefore can lead to formation of different end materials that can change the system's magnetic response [21].

This study reports on structural, morphological and magnetic properties of core-shell Au/Fe$_3$O$_4$ NPs, where high resolution transmission electron microscopy (HRTEM) technique has been used to assess the crystalline structure, morphology, and particle size distribution. Magnetization measurements at different temperatures were recorded to study the influence of particle size distribution, temperature dependence of the coercive field and occurrence of exchange bias effect. The latest being expected due to the interaction of the magnetically ordered region of the Fe$_3$O$_4$ shell with the disordered layers located in the inner and outer surface region of the Fe$_3$O$_4$ shell.

**Synthesis**

The Au/Fe$_3$O$_4$ CSNPs reported in the present study were synthesized by the thermal decomposition method adapting a procedure described elsewhere[14,20,22] and based on the mixing of Fe(III) acetylacetonate and Au(III) acetate in the presence of capping molecules. Firstly, for the formation of the Fe$_3$O$_4$ phase 1 mmol of Fe(III) acetylacetonate (Fe(acac)$_3$), 5 mmol of 1,2-hexadecanediol (reducing agent), a mixture of surfactants of 3 mmol oleylamine and 3 mmol oleic acid, and 10 mL of 1-octadecene (which has a boiling point of 315 °C) were mixed. The mixture was heated up to 120 °C in a three-neck round-bottom flask mounted on a temperature-controlled reflux system. Subsequently, 0.3 mmol of Au(III) acetate was added to the solution and the temperature was kept constant at 120 °C, for 30 min. Finally, the solution was heated up to 260 °C and this temperature was maintained constant for 150 min. We found that the time of 30 min was enough to complete the ligand exchange and the formation of core-shell Au/Fe$_3$O$_4$ nanostructures. The mixture was magnetically stirred under a flux of Ar gas during the whole process. After the synthesis (~ 30 min) the solution was cooled down to room temperature.

Surfactants were used during the synthesis in order to prevent particle agglomeration (and disordered growth of the NPs). Moreover, a gold precursor was carefully added and kept at 120 ºC in order to control the competitive growth process of different phases. The synthesis parameters were crucial to control the growth of the gold core phase at the beginning of the reaction, followed by formation of the magnetite shell phase. The core-shell nanoparticles are formed as the temperature increases and in the presence of surfactants. The obtained NPs were dispersed as a stable magnetic fluid (ferrofluid) and, in order to get a powdered sample, some amount was washed several times in a mixing solution of ethanol and hexane and finally dried at 70 °C during 24 h.

**Results and discussion**

Figure 1(A) shows the TEM images of the as-synthesized Au/Fe$_3$O$_4$ core-shell NPs. The high-contrast dark part in the core region of each particle has been assigned to the gold phase and the clear part surrounding the core regions has been assigned to the Fe$_3$O$_4$ phase. This difference in contrast is due to the difference in the electronic densities between Au and Fe$_3$O$_4$. TEM images have been used to assess the core size region (dark part) as well as the whole (dark + clear parts) regions (N≈ 900) and histograms for each region have been mounted using the Sturges method as observed in the inset of Figure 1(A). The histograms have been modeled by fitting the data to lognormal functions. The fitting parameters provided a mean core size of $<d_C>$ = 6.9 ±1.0 nm with dispersion index of $\sigma$ = 0.14 for the core region and a mean NP size of $<d_w>$ = 14.1±1.2 nm with dispersion index equals to $\sigma = 0.17$ for the whole particle size. Thus, a mean thickness of the shell region (Fe$_3$O$_4$) is estimated to be ≈3.5 nm. The HRTEM images of Au/Fe$_3$O$_4$ NPs were used to determine the inter-fringe space of the core and shell regions and the values we found matched the interplanar distances of the spinel structure of magnetite (space group Fd-3m) and cubic structure of gold (space group Fm-3m). Thus, Fast Fourier Transform (FFT) analysis of HRTEM images was carried out in a single particle, assessing the spot diffraction patterns of both Au and Fe$_3$O$_4$ phases (see the upper panel in the inset of Figure 1(b)). Data analysis indicates that some spots correspond to the (111), (022) and (113) atomic planes of the spinel Fe$_3$O$_4$ phase with interplanar distances of 4.84, 2.93 and 2.52 Å, respectively. Besides, some other spots correspond to the (111) and (022) atomic planes of the cubic phase of Au with interplanar distances of 2.35 Å and 1.46 Å, respectively. Those values are consistent with the standard pattern of the Joint Committee on Powder Diffraction Standards (JCPDS) for magnetite (card No. 75–449) and gold (card No. 89-3697) and confirm the formation of the Au-phase (core)/Fe$_3$O$_4$ (shell) structure determined from the TEM images. The HRTEM images clearly show the epitaxial growth of the Fe$_3$O$_4$ phase onto some facets of the core (gold) structure (see also Fig. S1). Figure 1(B) also shows the growth of the Au phase along the (111) and (022) atomic planes. The faceted nature of the core region and the epitaxial growth yielded a polycrystalline layer around the core region. The high angle annular dark field scanning transmission electron microscopy (HAADF-STEM) images performed on several Au/Fe$_3$O$_4$ NPs confirmed the core-shell morphology described above (see the inset at the bottom part of Figure 1(B)). Single-particle analysis for chemical composition was performed using the HAADF-STEM equipped with energy dispersive X-ray (EDX) spectroscope, confirming the Au/Fe$_3$O$_4$ core-shell formation.

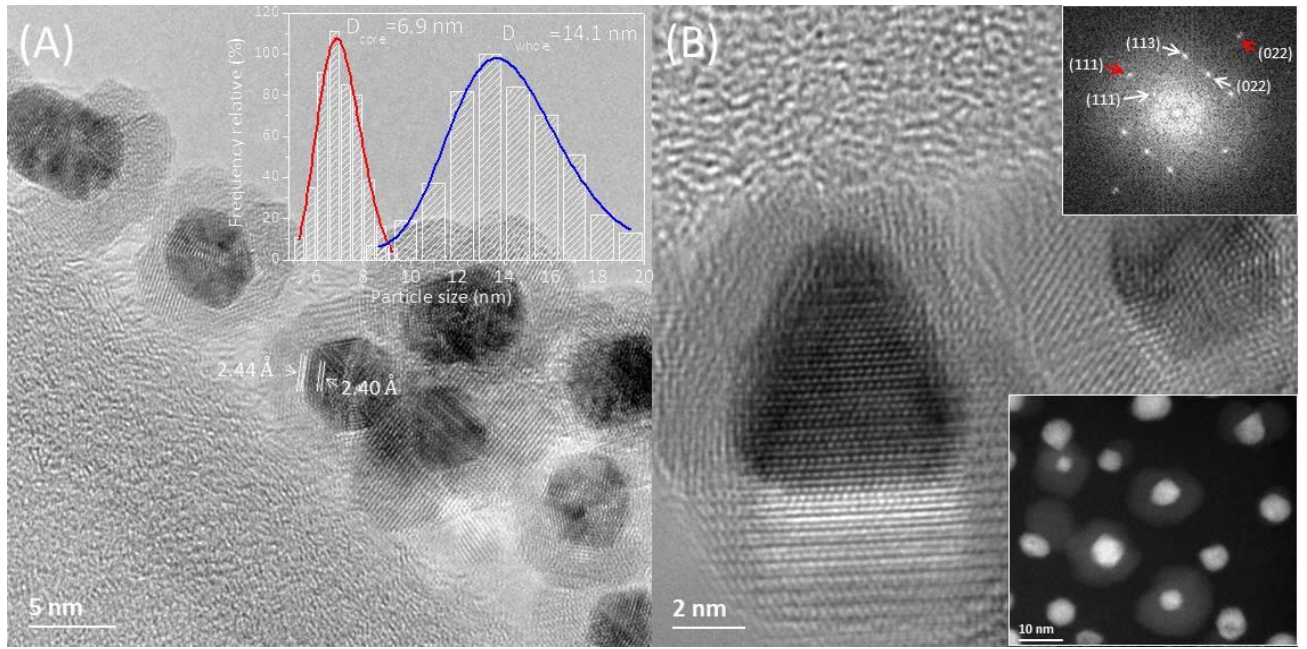

**Figure 1.** (a) TEM image of the Au/Fe$_3$O$_4$ CSNPs showing the successful formation of core-shell structure. In the inset it is shown the histogram of particle size mounted as described in the text. The solid lines represent the lognormal functions. (b) The HRTEM image of the NPs with its correspondent FFT at the top inset. In the low inset the HAADF-STEM image of the Au/Fe$_3$O$_4$ NPs is shown.

In order to investigate the magnetic properties of the Au/Fe$_3$O$_4$ NPs dispersed as ferrofluid field-cooled (FC) and zero-field-cooled (ZFC) magnetization curves were obtained under an applied magnetic field of 2.39 kA/m (see Figure 2(a)). The features showed by the ZFC-FC magnetization traces such as a maximum on the ZFC trace and irreversibility between both traces are consistent with the superparamagnetic (SPM) behavior. It is determined that the irreversibility starts well above (at ~190 K) the position of the maximum of the ZFC trace, around $T_{max}$~75 K, suggesting the occurrence of particle-particle interaction. Figure 2(c) shows the ZFC and FC traces obtained for the Au/Fe$_3$O$_4$ NPs after drying out the ferrofluid solvent (powder sample). As observed, in comparison with the ferrofluid sample the maximum of the ZFC curve is shifted to higher temperatures ($T_{max} \approx 195$ K). This result strongly suggests the enhancement of the magnetic interactions among the particles after the drying step, likely related to the formation of more aggregates of NPs while compared with the ferrofluid sample[23].

Based on a system composed of non-interacting single-domain particles with uniaxial anisotropy, Néel calculated the relaxation time using the relation $\tau = \tau_0 \exp(E_a/k_B T)$, where $\tau_0$ is the characteristic relaxation time (usually in the range of 10$^{-9}$-10$^{-13}$ s for SPM systems), $E_a$ is the energy barrier, $k_B$ is the Boltzmann's constant, and $T$ is the absolute temperature[24,25]. It is known that in a real system the effects of the particle size distribution, particle interaction, canted spins at the surface and magnetic anisotropy need to be taken into account for a better evaluation of the magnetic properties. Considering the effect of the particle size distribution, a system with a determined distribution of particle size gives rise to a distribution of blocking temperature $T_B$[26]. In this case, the zero-field cooled (ZFC) susceptibility can be expressed as:

$$\chi_{ZFC} = \frac{M_S^2}{3K_{eff}} \left[ \ln\left(\frac{\tau_m}{\tau_0}\right) \int_0^T \frac{T_B}{T} f(T_B) dT_B + \int_T^\infty f(T_B) dT_B \right] \quad (1)$$

whereas the field cooled (FC) susceptibility reads:

$$\chi_{FC} = \frac{M_S^2}{3K_{eff}} \ln\left(\frac{\tau_m}{\tau_0}\right) \left[ \frac{1}{T}\int_0^T T_B f(T_B) dT_B + \int_T^\infty f(T_B) dT_B \right]. \quad (2)$$

The first term in both equations is the contribution of particles in the superparamagnetic state; meanwhile, the second term refers to particles in the blocked state. Here, $M_S$ is the saturation magnetization, $K$ is the uniaxial anisotropy constant, $f(T_B)$ is the distribution function of blocking temperature and $\tau_m$ is the measuring time window ($\tau_m = 10^2 s$, for DC magnetic measurements)[27].

Using the ZFC and FC curves one can obtain the blocking temperature distribution, which is given by: $f(T_B) \sim -d(\chi_{FC} - \chi_{ZFC})/dT$ [28]. As observed in Figure 2(b) the experimental data shows bimodal features and suggests the occurrence of two blocking temperature distributions, which can be modeled using lognormal distributions of blocking temperatures:

$$f(T_B) = \frac{A}{\sqrt{2\pi}\,\sigma_1 T_B} exp\left[-\frac{1}{2\sigma_1} ln^2\left(\frac{T_B}{\langle T_{1B} \rangle}\right)\right] + \frac{(1-A)}{\sqrt{2\pi}\,\sigma_2 T_B} exp\left[-\frac{1}{2\sigma_2} ln^2\left(\frac{T_B}{\langle T_{2B} \rangle}\right)\right],$$

where $\langle T_{iB} \rangle$ is the mean value of the blocking temperature, $A$ is a weight factor, and $\sigma_i$ is the polydispersion parameter[27]. The data fit is shown in Figure 2(b) and the obtained values are $\langle T_{1B} \rangle = 59$ K, $\langle T_{2B} \rangle = 20$ K, $\sigma_1 = 0.48$, $\sigma_2 = 0.63$ and $A=0.68$ for the Au/Fe$_3$O$_4$ ferrofluid sample. Those values obtained from the fit were used to simulate the ZFC and FC curves using Eqs. (1) and (2) and the results are shown in Figure 2(a). As observed, the ZFC and FC curves of the Au/Fe$_3$O$_4$ ferrofluid sample are well reproduced using $K_{eff} = 1 \times 10^4$ J/m$^3$ and $M_s = 11.4$ $Am^2/kg$. The $M_s$ value is in agreement with the measured value whereas the $K_{eff}$ value is in good agreement with the reported value for bulk magnetite[29]. Differences between the experimental and calculated FC curves could be assigned to the occurrence of particle-particle interaction, once the effect of those interactions is the flattening of the FC curves, in agreement with the literature[30]. In the low temperature region the blocking temperature distribution has been assigned mainly to the contribution of non-interacting magnetite NPs. The occurrence of a blocking temperature assigned to individual magnetite NPs formed in the Ag/Fe$_3$O$_4$ nanostructure has been reported in the literature[31]. Also, at low temperature the second distribution of blocking temperatures is consistent with the small shoulder observed in the ZFC curve recorded from the dried sample (see Figure 2(c)).

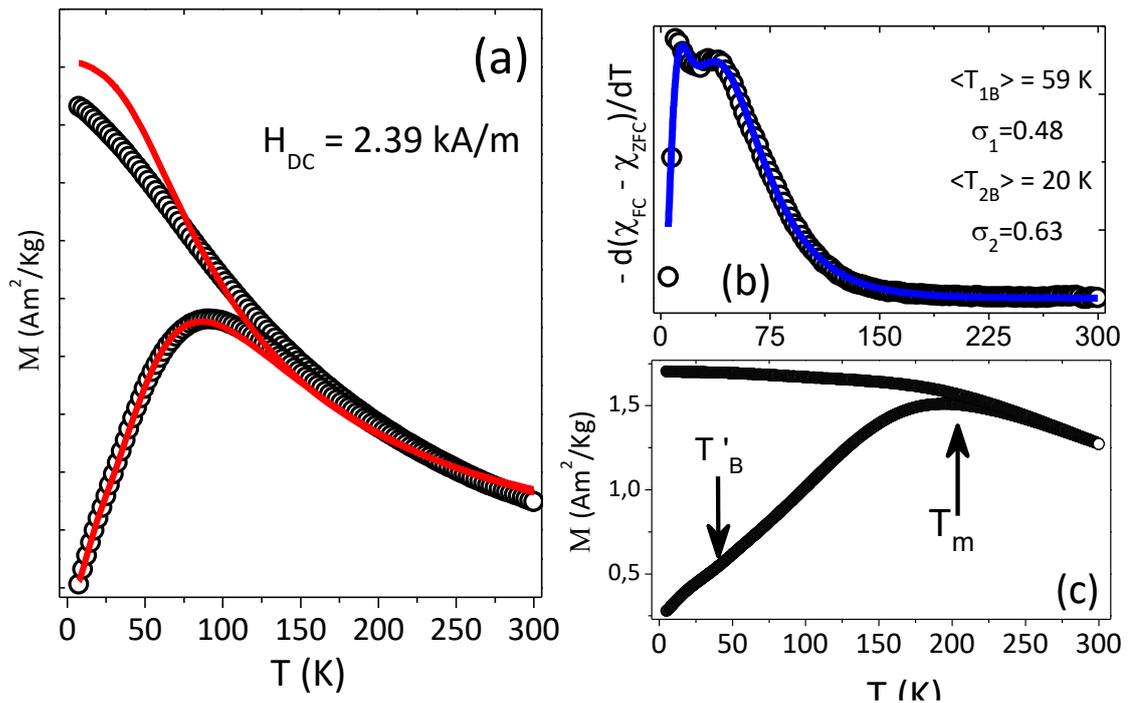

**Figure 2.** (a) ZFC – FC magnetization curves ($H$ = 2.39 kA/m) of Au/Fe$_3$O$_4$ NPs (ferrofluid sample). (b) $-d(\chi_{FC} - \chi_{ZFC})/dT$ as a function of the temperature. The solid line represents the lognormal function as a function of the temperature used to fit the data. (c) ZFC – FC traces obtained for the dried sample.

In order to probe the effects of particle size distribution and/or particle-particle interaction, we have performed systematic magnetization measurements as a function of applied field, *M* vs *H*, at several temperatures. Figure 3 (a) shows the *M* vs *H* curves obtained at 5 K and 300 K for the ferrofluid sample. As can be inferred, the *M* vs. *H* curve at 300 K shows no hysteresis and a rapid magnetization increase in the region of low magnetic fields, which is consistent with a superparamagnetic behavior, in agreement with the temperature dependence of the ZFC and FC curves previously discussed. The saturation magnetization found for the dried sample at *T* = 300 K showed a value $M_S$ = 28 Am$^2$/kg$_{Fe3O4}$, in agreement with previously observed values in gold/iron oxide, core/hollow-shell NPs[8]. The reduction of $M_S$ with respect to the bulk magnetite values can be assigned to the surface spin disorder, also reflecting in the non-saturating behavior of the *M* vs. *H* curves at high applied magnetic fields (up to 7 T).

The temperature dependence of the coercive field ($H_C$) is shown in Figure 3 (a), with the hysteresis loops obtained at 5 and 300 K shown in the inset. As observed in the main panel, *Hc* rapidly decreases as the temperature is increased and above ~60 K the value is very small and tends to vanish, confirming that ensembles of CSNPs are in the superparamagnetic (SPM) state above that temperature, in agreement with the ZFC-FC traces analysis (see Figure 2). However, the temperature dependence of the coercive field can be attributed to the particle size distribution and/or interaction effects. Nunes et al. proposed an alternative model which takes into account the effect of the particle size distribution on the total coercive field in which case the coexistence of both blocked and superparamagnetic states is considered[27]. In this case, the temperature dependence of the coercive field is given by:

$$\langle H_C \rangle_T = \frac{M_r(T)}{\chi_S + \frac{M_r(T)}{H_{CB}(T)}}, \qquad (3)$$

where $M_r(T)$ is the remanence magnetization which depends on the temperature, $\chi_S(T)$ is the susceptibility of the superparamagnetic particles at a given temperature *T*, and $H_{CB}(T)$ is the coercive field of the blocked particles. The remanence magnetization is related to $T_B$, according to the following relation:

$$M_r(T) = \alpha M_s \int_T^\infty f(T_B) dT_B,$$

where $\alpha$ is a parameter that depends on the degree of magnetic anisotropy of the NP[32]. The superparamagnetic susceptibility, $\chi_S$, for a system with a distribution of particle size can be calculated using[33]:

$$\chi_S = \frac{25 M_S}{3\, K_{eff}\, T} \int_0^T T_B f(T_B) dT_B.$$

It is known that for a system with a particle size distribution one correlates it to a distribution of $T_B$. Taking into account only blocked particles at $T < T_B$, the coercive field is given by:

$$H_{CB}(T) = \alpha\, \frac{2\, K_{eff}}{M_S} \left[ 1 - \sqrt{\frac{T}{\langle T_B \rangle_T}} \right],$$

where $\langle T_B \rangle_T$ takes into account only the volume fraction of blocked particles at temperature *T*,

$$\text{defined as } \langle T_B \rangle_T = \frac{\int_T^\infty T_B f(T_B) dT_B}{\int_T^\infty f(T_B) dT_B}.$$

Using the values obtained from the ZFC-FC analyses, the temperature dependence of the coercive field is well modeled by the Eq. (3), as shown in Figure 3(a). Nunes et al.[27] have proposed that the effect of random interactions can be pictured by the parameter $\gamma$, which is an empirical parameter used in $f(\gamma T_B)$. In our case, this parameter was found to be $\gamma = 1.28$, which could be considered small and indicates weak interactions. It is determined that the contribution of superparamagnetic particles and the mean blocking temperature are important to describe the coercive field behavior in a wide temperature range.

The possible occurrence of exchange anisotropy or exchange bias in the Au/Fe$_3$O$_4$ core-shell NPs was also investigated. The magnetic hysteresis loops of the Au/Fe$_3$O$_4$ NPs (ferrofluid sample) were obtained in field cooled condition ($H_{FC}$ = 2 T) at different temperatures from 5 to 300 K. The exchange anisotropy field ($H_{EX}$) has been determined from the loop shift along the field axis. The temperature dependence of the $H_{EX}$ is depicted in the main panel of Figure 3(b) while the hysteresis curve obtained at 5 K is shown in the inset. In contrast to the $H_C$, the $H_{EX}$ shows a rapid increase below ~40 K. For temperatures above ~40 K, the absence of exchange bias effect is determined; meanwhile, $H_c$ shows non-zero values above this temperature. This result suggests the occurrence of a spin freezing phenomenon below ~40 K that we speculatively assigned to the disordered spins located likely at the two surfaces of the magnetite shell: in the inner surface of the Au/Fe$_3$O$_4$ interface and in the outer surface of the magnetite shell. A schematic representation of the different magnetic regions is shown in Figure 3(c). The onset of the magnetic disorder is related to symmetry breaking and broken bonds happening at the interface and particle surface, which drives the system to magnetic frustrations. Due to the action of the applied magnetic field the ensemble of spins of all regions are aligned along the magnetic field direction at high temperatures. When the system is cooled down below the freezing temperature (~40 K) the frozen spins act as pinning agents during the reversal of spins of the ordered region, driving to the shift of the hysteresis loop. The occurrence of the exchange anisotropy field related to a surface magnetic spin layer disorder has been previously reported in $\gamma$-Fe$_2$O$_3$, CoO/Fe$_3$O$_4$ and in hollow Fe$_3$O$_4$ NPs. The stronger exchange bias effect observed in the latter system in comparison to the one investigated in this study could be related to the larger magnetic disorder layer at the innermost and outermost surfaces of the magnetic shell[10,18,34,35,36].

The random-field model (RFM) of exchange anisotropy developed for ferro/antiferromagnetic multilayers has been used[37] to evaluate the thickness of the disordered spin layer in $\gamma$-Fe$_2$O$_3$ nanoparticles[9], but the thermal dependence of $H_{Ex}$ has not been considered in that report. Within the RFM framework, the exchange anisotropy field $H_{Ex}$ is related to the magnetic anisotropy by: $H_{EX} = \Lambda K_{eff}^{1/2}$, where $\Lambda$ is a parameter related to the effective magnetic anisotropy ($K_{eff}$). The temperature dependence of $K_{eff}$ can be modeled using the empirical Brukhatov-Kirensky relation[38], $K_1(T) = K_1(0)\exp[-\alpha T^2]$. Then, the thermal dependence of $H_{EX}$ is given by:

$$H_{EX}(T) = \Lambda K_{eff}^{\frac{1}{2}} = H_{EX}(0)e^{-BT}, \qquad (4)$$

where $H_{ex}(0)=\Lambda K_1(0)$ is the exchange bias field at $T=0$ K and $B$ is a constant. Using the above expression the experimental data are well modeled when the $H_{EX}(0)$ = 7.2 kA/m and $B$= 1.3 10$^{-2}$ K$^{-1}$.

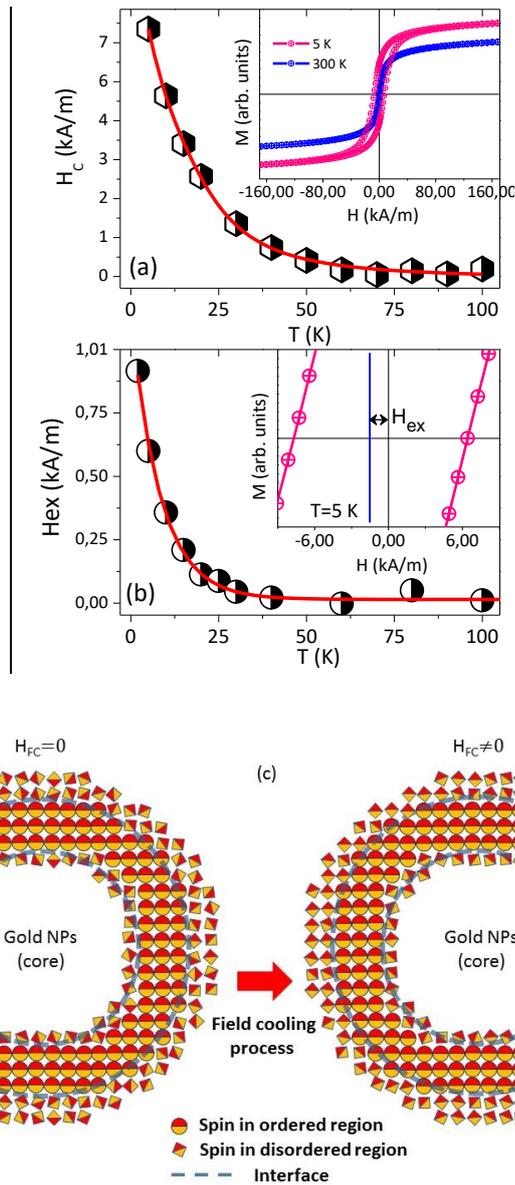

**Figure 3** (a) Temperature dependence of the coercive field of the Au/Fe$_3$O$_4$ NPs fluid sample. Here, the temperature dependence of the coercivity was fitted for using the generalized model that proposes a temperature dependence of blocking temperature due to the coexistence of blocked and unblocked particles. The inset shows the hysteresis loops of the Au-Fe$_3$O$_4$ NPs obtained at 5 K and 300 K. (b) The exchanged bias field as a function of temperature obtained after a FC process with a field of 2 T. The inset illustrates the magnetization hysteresis loop shift at 5 K. (c) Schematic representation of the different magnetic regions in a particle and the effect produced by the cooling in a magnetic field.

In order to analyze the effect of interactions we have studied the dynamical properties through the *T*-dependence of the in phase ($\chi'(T)$) and out-of-phase ($\chi''(T)$) components of the AC magnetic susceptibility, in an oscillating magnetic field of 5 Oe and varying the excitation frequency in the range 15 Hz $\leq f \leq$ 1 kHz. As shown in Figure 4(a), both $\chi'(T)$ and $\chi''(T)$ components show a peak. The position of the maximum $T_m$ observed in the $\chi'$ vs. T curve is shifted to higher temperatures as the frequency is increased. Also, the position of $T_m$ is consistent with the position of the maximum determined from ZFC and FC curves. At low temperatures, a broad shoulder is distinguishable in the $\chi'$ vs. *T* curve. Further evidence of the presence of the second peak at ~30 K has been determined from the $d\chi''/dT$ vs. *T* curve (see Figure S2, in supplementary material). The presence of the second peak is also consistent with the results determined from the ZFC and FC measurements, which was assigned to the magnetic response of non-interacting magnetite NPs[39].

It is known that the dynamical response of an ensemble of fine particles is determined by the measuring time $\tau_m \sim 1/\omega$ of each experimental technique[40]. Within the Néel theory, the relaxation of the magnetic moment of a single-domain particle over the anisotropy energy barrier exhibits an exponential dependence on the temperature given by: $\tau = \tau_0 \exp(E_a/k_B T)$. Using the Néel relation to analyze the frequency dependence of $T_m$, the fit of the experimental data (see Figure 4(b)) provides an activation energy of $E_a/k_B = 1231$ K and a characteristic relaxation time of $\tau_0 = 5 \times 10^{-11}$ s has been obtained. The $\tau_0$ value is in the range expected for a system showing superparamagnetic behavior. Moreover, the activation energy barrier can be given by:[41]

$$E_a = E_{ani} + E_H + E_{ex} + E_{d-d}, \quad (5)$$

where $E_{ani}$ is the anisotropy energy, $E_H$ is the Zeeman energy, $E_{ex}$ is the exchange energy in the system and $E_{d-d}$ is the dipolar interaction energy. Under low external magnetic field, the Zeeman term is $E_H \approx 0$. As the particles show in average a certain edge-to-edge spacing, the exchange interactions can be assumed negligible[42]. The remaining terms are the anisotropy and dipolar interactions. Assuming that the dipolar term is negligible the activation energy barrier reads: $E_a = E_{ani} = K_{eff} V$. Using the mean thickness of the Fe$_3$O$_4$ shell, we estimated an effective anisotropy value of $K_{eff} = 1.3 \times 10^4$ J/m$^3$. We found the estimated effective anisotropy value is in the range of values reported for the first-order magnetocrystalline anisotropy constant of bulk magnetite (1.1-1.3×10$^4$ J/m$^3$)[43]. This result indicates that the particle-particle interactions are weak in the studied sample. In fact, a rough analysis of the dipolar interaction can be done by considering a pair of particles whose dipolar interaction energy is given by: $E_{d-d} = \mu^2/r^3$, where $\mu$ is the magnetic moment of a particle and $r$ is the center-to-center distance between them. Taking into account the thickness of the magnetic shell assessed from TEM analysis an energy of $T' = E_{d-d}/k_B \sim 980$ K is estimated for particles which are in close contact to each other. Once the particles show a physical separation between them due to the existing surfactant layer, a fast decrease in the strength of the dipolar interaction energy is expected according to $\frac{E_{d-d}(\alpha)}{E_{d-d}(0)} \sim d_W^3/(d_W + \alpha)^3$, where $\alpha$ is the edge-to-edge spacing and $d_W$ is the mean particle size. For a $\alpha/d_W \sim 2$, the particle-particle interaction strength becomes $\sim 3\%$ of $E_{d-d}(0)$. This indicates that the dipolar interaction energy is weaker than the anisotropy energy, leading to an ensemble of weakly interacting particles in the studied sample [42,44].

The latter result is confirmed by analyzing the frequency dependence of the freezing/blocking process using the empirical parameter: $\Phi = \frac{\Delta T_m}{T_m \Delta \log_{10}(f)}$, where $\Delta T_m$ is a shift of $T_m$ in the $\Delta \log_{10}(f)$ interval. We obtained a value of $\Phi=0.098$ for the Au/Fe$_3$O$_4$ NPs which is consistent with values previously reported for weakly interacting systems ($\Phi = 0.1$-$0.13$)[45,46], reinforcing the already mentioned weak contribution of dipolar interaction among the particles.

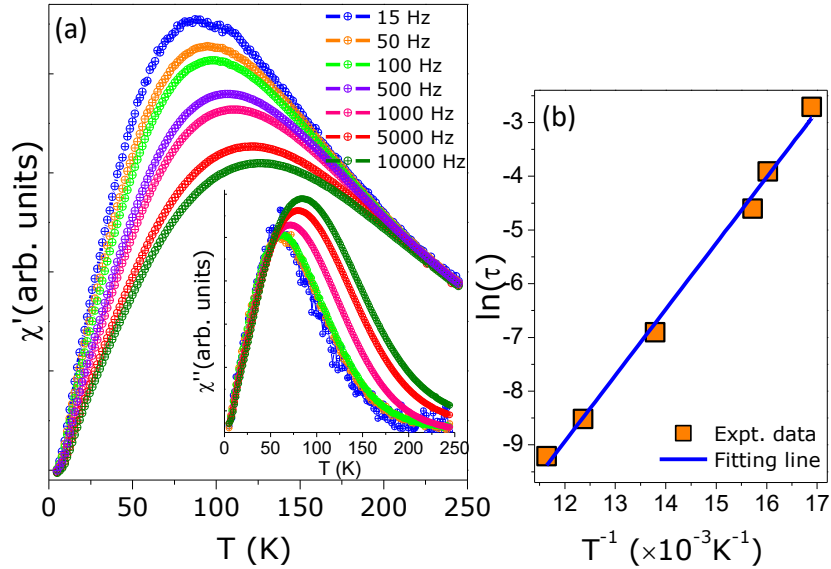

**Figure 4** (a) In-phase component of the ac susceptibility, $\chi'(T)$, as a function of the temperature obtained by varying the excitation frequency and with an oscillating field of 5 Oe field for the ferrofluid sample containing $Au/Fe_3O_4$ CSNPs. The inset shows the out-of-phase component, $\chi''(T) vs. T$. (b) The relaxation time as a function of the inverse of the maximum ($T_m$) determined from the $\chi'(T) vs. T$ curve. The solid line is the fit to Néel-Arrhenius relation.

## Conclusions

We successfully synthesized core-shell $Au/Fe_3O_4$ NPs using a thermal decomposition synthesis route, where the core is found to be Au and the shell $Fe_3O_4$. A mean particle size of the core-shell structure has been determined ~14 nm with a $Fe_3O_4$ shell thickness of ~3.5 nm, as estimated from the electron microscopic micrographs. Magnetic data analysis suggests weakly interacting NPs. It has been determined the occurrence exchange anisotropy effect that is originated from the interaction of a magnetically ordered layer in the magnetite shell and disordered spins located at the interface $Au/Fe_3O_4$ and external surface of the magnetite shell. The exchange bias field shows a decreasing tendency with the increase of the temperature and disappears at ~40 K, which is below the blocking temperature of the $Au/Fe_3O_4$ NPs. This result is in agreement with the picture of freezing of spins located in the magnetically disordered regions.

## Methods

### Instruments and characterization

The morphology and structural information of core-shell nanoparticles was analyzed by high resolution transmission electron microscopic (HR-TEM). The images were obtained by using a FEI Tecnai F30 microscope operated at an acceleration voltage of 300 kV. The microscope was equipped with a HAADF (high angle annular dark field) detector for STEM mode and EDX (X-ray energy disperse spectrometry). Lattice fringes were measured from the fast-Fourier transform of HRTEM images, using Gatan Digital Micrograph software.

The magnetic properties were studied by carrying out magnetization measurements in a superconducting quantum Interference device (SQUID), model MPMS3 (Quantum Design, Inc., San Diego, CA) in a wide range of temperatures (from 2 to 300 K) and applying magnetic fields up to 7 T. AC magnetic susceptibility measurements were performed by using an AC susceptometer, ACMS module of a physical property measurement system (PPMS, Quantum Design, Inc.).

**Acknowledgements** This work was supported by the Brazilian agencies CNPq and CAPES (BEX 6932/15-0) through a financial support in the PhD Student Exchange Program performed in the Institute of Nanoscience of Aragón. The work in Spain was also supported by the Spanish Ministerio de Economia y Competitividad (MINECO) through project MAT2013-42551 and the Aragon Regional Government (DGA, Project No. E26) Technical support from LMAINA and SAI-UZ is acknowledged.

**Author contributions**
L. León Félix synthesized the nanoparticles, designed and carried out the experiments and did the analysis of the results. M.A.R. Martinez and M.H. Sousa helped in the sintering of the nanoparticles and the experimentation. J.A.H Coaquira and J. Mantilla performed the magnetic measurements and analyzed the results. G.F. Goya performed the structural characterization of the samples. L. de los Santos Valladares and C.H.W. Barnes wrote the manuscript and contributed to the structural and magnetic analysis. All authors contributed in the analysis and discussion of the results.

**Competing financial interest:** The authors declare not competing financial interest.